\documentclass[pra,twocolumn,amsmath,amssymb]{revtex4-2}
\usepackage{bm}% bold math

\usepackage{epsfig}
\usepackage{xcolor}

\begin{document}

\title{Phase-amplitude separation of wave function as local gauge transformation}  
\author{A.\  R.\  P.\  Rau\footnote{ravirau@lsu.edu} \\
Department of Physics and Astronomy, 
Louisiana State University, 
Baton Rouge, Louisiana 70803}

\begin{abstract}
A quantum-mechanical wave function is complex, but all observations are real, expressible through expectation values and transition matrix elements that involve the wave functions. It can be useful to separate at the outset the amplitude and phase as real quantities that together carry the same information that is contained in the complex wave function. Two main avenues for doing so go way back in the history of the subject and have been used both for scattering and bound states. A connection is made here to gauge transformations of electrodynamics where the advent of quantum mechanics and later quantum field theory showed the central role that local gauge transformations play in physics.

\end{abstract}

%\pacs{03.65.Ge, 03.65.Sq, 31.15.xj}

\maketitle
    
\section{Introduction}

We consider the radial Schr\"{o}dinger equation of a three-dimensional non-relativistic system

\begin{equation}
[d^2/dr^2 + k^2(r)] u(r) =0, \, k^2(r) = E -V(r),
\label{eqn1}
\end{equation}
where $u(r)$ is the ``reduced" radial function, $u(r) = r R(r)$, so as to remove the linear $d/dr$ in the kinetic energy (as a result, $u(0)=0$), $k(r)$ is the local wave-number (or local linear momentum $p(r) =\hbar k(r)$), $E$ the total energy, and $V(r)$ the potential energy. The $V(r)$ can include the so-called angular momentum barrier for a non-zero angular momentum $\ell$ besides any external potential, long or short range. Both $E-V$ and the total energy may be of either sign depending on the asymptotic energy of interest and can embrace bound and scattering states as well as resonances in a multi-channel context. 

General solutions of Eq.(1) are superpositions of sines and cosines of standing waves or the pair of complex exponentials describing traveling waves of $k(r)$ (equivalently, local wavelength $2\pi/k(r)$). Two standard methods of writing $u(r)$ as an amplitude and a phase function, called phase-amplitude methods (PAM), will be considered separately below. The first is naturally adapted to the well-known Jeffreys-Wentzel-Kramers-Brillouin (JWKB) method or approximation which proves powerful in quasi-classical situations and is analogous to the eikonal method familiar in physical optics in a near geometrical-optics situation. The second is useful in scattering theory in using an asymptotic phaseshift and a companion amplitude or intensity (upon squaring), that phaseshift immediately connected to observable scattering cross-sections or energy and other parameters of bound and resonance states.

In a long and distinguished career, Chris Greene has made many important contributions to physics. In one subject, quantum defect theory \cite{ref1, ref2, ref3}, his contributions go back to his graduate student days. This subject considers complex atomic and molecular spectra, bound and scattering states as well as resonances in a unified framework with a pair of so-called regular and irregular solutions of the stationary state Schr\"{o}dinger equation with boundary conditions at the origin and at infinity. Chris was involved in developing these pairs for a wide range of potentials, long and short range, adapting the JWKB method for this purpose, and then in applying them to complex systems, separating long and short range contributions and parametrizing them in a small set of quantities \cite{ref4, ref5, ref6}. Adopting the Wigner R-matrix philosophy \cite{ref7}, the small set of short range parameters that are insensitive to the asymptotic energy can then be obtained either from experiment or variational calculations of the R-matrix \cite{ref8}. This note that provides my own perspective on PAM, variational principles, and their interpretation as gauge transformations that are central to quantum field theories is offered in admiration of Chris Greene's contributions over the years and in friendship.

\section{Milne-Young-Wheeler Method}

When $k(r)$ is a constant, $k$, solutions of Eq.(1) are immediate, a single sine or cosine or complex exponential in $kr$. Normalization per unit energy of the wave function requires a multiplicative factor $k^{-1/2}$ \cite{ref2}. The JWKB method rests on arguing for slowly varying $k(r)$ (when $V(r)$ varies slowly over the wavelength) that an approximate solution is, therefore,

\begin{equation}
u_{JWKB} (r) = k^{-1/2}(r) \exp [\pm \, i \int^r k(r') dr'].
\label{eqn2}
\end{equation}

This is an approximate solution. Inserting into Eq.(1) leaves higher derivatives of $k(r)$ which are neglected in the JWKB treatment as of higher order. This statement can be alternatively rendered as saying that  Eq.(2) actually satisfies, as is easily verified,

\begin{equation}
[\frac{d^2}{dr^2} + k^2(r)] u_{JWKB}(r) = [\sqrt{k(r)} \frac{d^2}{dr^2} k^{-1/2}(r)]u_{JWKB}(r).
\label{eqn3}
\end{equation}
The higher derivatives of the local wavelength are gathered on the right-hand side.

The method of Milne-Young-Wheeler \cite{ref2, ref9, ref10, ref11} takes the cue from this to write the exact solution of Eq.(1) in a form suggestive of JWKB,

\begin{equation}
u (r) = K^{-1/2}(r) \exp [\pm \, i \int^r K(r') dr'],
\label{eqn4}
\end{equation}
with an unknown $K(r)$, the equation for it given by inverting the argument that took Eq.(2) to Eq.(3), namely,

\begin{equation}
K^2(r) - [\sqrt{K(r)} \frac{d^2}{dr^2} K^{-1/2}(r)] = k^2(r).
\label{eqn5}
\end{equation}

This is a nonlinear equation in the function $K(r)$, not surprising because the writing of $u(r)$ in exponential form is a nonlinear expression. Nonlinearities are characteristic of any such recasting of the linear Schr\"{o}dinger differential equation in JWKB or Hamilton-Jacobi form. $K(r)$ occurs in two places in Eq.(4), as an amplitude $\alpha(r)$ and a phase $\phi(r)$, with

\begin{equation}
\alpha(r) = K^{-1/2}(r), \,\, d\phi/dr = K(r).
\label{eqn6}
\end{equation}

Cast in terms of the amplitude function, the Milne-Young-Wheeler Eq.(5) takes the form

\begin{equation}
[d^2/dr^2 + k^2(r)] \alpha (r) = 1/\alpha^3(r).
\label{eqn7}
\end{equation}

Solving this single nonlinear equation self-contained in $\alpha(r)$ is equivalent to solving the linear Schr\"{o}dinger Eq.(1) and, once solved, $\phi(r)$ follows by a single quadrature integral of $\alpha^{-2}(r)$ from Eq.(6), yielding the full $u(r)$. This is to be contrasted with the second PAM we will consider in Sec.III with the opposite structure, of a self-contained nonlinear equation for a phase function, and then quadrature from it giving the amplitude function.

Also notable is that the above procedure provides a convenient way \cite{ref2, ref4, ref5} of obtaining the pair of linearly independent solutions of Eq.(1), often called $(f, g)$, $f$ the so-called regular and $g$ the irregular solution, of a second-order differential equation for any $V(r)$. This has been extensively used in the subject of quantum defect theory, the JWKB approach providing pairs of solutions for many long and short range potentials as well as convenient parameters for expressing them that are of use in analyzing complex atomic and molecular spectra \cite{ref2, ref12}. One writes

\begin{equation}
f(r) = \sqrt{\frac{2}{\pi}} \alpha(r) \sin \phi (r), \,\, g(r) =- \sqrt{\frac{2}{\pi}} \alpha(r) \cos \phi (r), 
\label{eqn8}
\end{equation}
the otherwise inconsequential numerical factors chosen to fix the Wronskian, $W(f, g) = 2/\pi$ and the signs for the conventional lag by $\pi/2$ of the irregular's phase relative to the regular's \cite{ref2}. 

\section{Dashen-Babikov-Calogero Method}

An alternative method \cite{ref13, ref14, ref15} for separating amplitude and phase of $u(r)$ in Eq.(1) is to again start as in Eq.(8) by writing a solution $f(r) = \sqrt{\frac{2}{\pi}} \alpha(r) \sin \phi (r)$. However, the redundancy in replacing a single $u$ by two functions is removed by a constraining form for the derivative, $df(r)/dr = \sqrt{\frac{2}{\pi}} \alpha(r) k(r) \cos \phi (r)$, which leads to the set of equations

\begin{eqnarray}
d\phi(r)/dr &=& k(r) + (k'(r)/2k(r)) \sin 2\phi (r), \\
d\alpha(r)/dr &=&- (k'(r)/k(r)) \alpha(r) \cos^2 \phi (r). 
%\label{eqn9}
\end{eqnarray}

Note that it is now $\phi(r)$ that obeys a stand-alone nonlinear (much higher nonlinearity than in the Sec.II procedure) equation. Once solved, it takes but a simple quadrature over it to get the amplitude $\alpha(r)$.

A special form of the above general procedure, treated extensively in a book on scattering theory devoted to PAM methods, focuses on separating $V(r)$ into long and short range components (denoted $v^{(s)}(r)$), and writing 

\begin{equation}
F(r) = \sqrt{\frac{2}{\pi}} \alpha(r) [f(r) \cos \delta^{(s)} (r)-g(r) \sin \delta^{(s)}(r)]. 
%\label{eqn9}
\end{equation}
With $(f, g)$, the pair in Eq.(8), describing the long range forces in the problem, and the conventions mentioned at the end of Sec.II, this $F(r)$ asymptotically is just as in $f(r)$ but only in having the short-range phaseshift in addition.

With the natural choice $\delta^{(s)} (0) =0$, the equations for phase $\delta^{(s)} (r)$ and amplitude $\alpha(r)$ are

\begin{eqnarray}
\frac{d\delta^{(s)}(r)}{dr}\! &=&\! -\frac{2}{W(f, g)} v^{(s)} [f \cos \delta^{(s)}(r)\!\! -\! g \sin \delta^{(s)} (r)]^2 \\
\frac{d\alpha(r)}{dr} &=& \frac{2}{W(f, g)} \alpha(r) v^{(s)} [f \cos \delta^{(s)}(r)\!\! - \!g \sin \delta^{(s)} (r)]  \nonumber \\
                     & \times & [f \sin \delta^{(s)}(r) + g \sin \delta^{(s)} (r)]. 
%\label{eqn9}
\end{eqnarray}

The first-order differential equation in Eq.(13) is really in $\ln \alpha(r)$ and that function is obtained by immediate integration once $\delta^{(s)}$ is determined from Eq.(12). Consistent with our choices in Eq.(8) and continuum normalization, the boundary condition on Eq.(13) is that $\alpha (\infty) =1$ and all $\alpha (r)$ are measured relative to it. These ``switches" between the two functions, $\delta^{(s)}$ and $\alpha$, that they are fixed by conditions at origin and $\infty$, respectively, the switch in sign on the right-hand side of Eqs.(12,13), and that the right-hand side of Eq.(13) is the functional derivative with respect to $\delta^{(s)}$ of the corresponding side in Eq.(12) will get further illumination in a variational formalism in Sec.IV that we turn to next.

\section{Phase and Amplitude functions as adjoints in a variational formalism}

The PAM method in Sec.III gives one non-trivial first-order differential equation for the phase function in Eq.(9) or Eq.(12). But the equations are nonlinear and may be difficult to solve for certain potentials and resultant $k(r)$. Applying a general method \cite{ref16} for constructing variational principles to Eq.(12), to determine the desired phaseshift at $r\rightarrow \infty$, $\delta^{(s)} (\infty)$, we start with a ``trial" (subscript $t$) solution $\delta_t^{(s)} (r)$ which is presumed known at all $r$. It may contain parameters that may be varied as usual in any application of variational principles to get the best final result. All $\delta_t$ will be chosen to satisfy the boundary condition at the origin, $\delta_t(0)=0$. From now on, for simplicity, we will drop the superscript $(s)$ and argument $r$ of $\delta_t$ and absorb the factor $2/W(f, g)$ into $v^{(s)}$ Eqs.(12,13). 

Using such a $\delta_t$, a function known everywhere at all $r$, we write

\begin{equation}
\delta_v (\infty)= \delta_t(\infty) - \int_0^{\infty} \!\! dr L(r) [ \frac{d\delta_t}{dr} +v^{(s)}(f \cos \delta_t-g \sin \delta_t)^2] 
%\label{eqn9}
\end{equation}             
as the variational estimate $\delta_v (\infty)$ where we have introduced a so-called ``Lagrange" function $L(r)$, as yet undetermined and undefined. The logic of this general construction \cite{ref16} is that were the trial solution to be the exact $\delta$ (a function of all $r$), we of course want its value at $\infty$, the exact phaseshift. And indeed, whatever $L(r)$, Eq.(14) would give this result because its factor in square brackets vanishes, at all $r$, by virtue of Eq.(12). 

Of course, an arbitrarily guessed or trial function $\delta_t(r)$ and its value $\delta_t(\infty)$ is unlikely to be so and will differ from the exact function and value in ``first order" $d\delta \equiv \delta_t -\delta$. Now, we cancel all such first-order terms in $d\delta$ in Eq.(14), doing an integration by parts of the first term in derivative $d/dr$. This transfers the derivative to act on $L(r)$ (with multiplicative $d\delta$) and introduces the value of $Ld\delta$ at the two limits of integration, 0 and $\infty$. Since all $\delta_t$ are chosen to satisfy the boundary condition of being zero at the origin, all $d\delta$ vanish at this limit and the only surviving contribution is of $Ld\delta$ at $\infty$. Both the exact $\delta$ and $d\delta$ are unknown at $\infty$ but the latter can be made to cancel the similar term in the first term on the right-hand side of Eq.(14) by choosing the condition $L(\infty)=1$. The remaining terms multiplying $d\delta$ under the integral are also set to zero by setting its integrand to zero at all $r$. This now defines the function $L(r)$ through a first-order differential equation and that boundary condition $L(\infty)=1$. And we now recognize that this is the same as Eq.(13) upon setting $L(r) = \alpha^{-2}(r)$. 

This now casts light on Sec.III, that the functions $\delta^{(s)}(r)$ and $\alpha^{-2}(r)$ are variational ``adjoint" functions \cite{ref16}. Such general constructions through incorporating the original defining functions such as Eq.(12) as a constraint through a Lagrange function (always introduced linearly) and then obtaining the defining equations for those Lagrange adjoints to make the expression such as Eq.(14) have no first order errors switches the role of the limits. The original function is given by a boundary condition at the origin, its value at $\infty$ what is sought, but the Lagrange adjoint function has its condition set at the opposite end, at $\infty$, and its value at the origin $L(0)$ contains the desired physics. The squared wave function at the origin, which is indeed the amplitude function at the origin is of considerable significance \cite{ref2}.

Gathering together the above results on $L$ and $\delta_t$, if such a trial $\delta_t ^{(s)}(r)$ as a solution of Eq.(12) and the $\alpha_t(r)$ that follows immediately from Eq.(13) are plugged into Eq.(14), namely,

\begin{equation}
\delta_v (\infty)\!=\! \delta_t(\infty) - \int_0^{\infty}\!\!\! dr \alpha_t^{-2}(r) [ \frac{d\delta_t}{dr} +v^{(s)}(f \cos \delta_t-g \sin \delta_t)^2], 
%\label{eqn9}
\end{equation}        
will give a variationally accurate value, presumably better than the $\delta_t(\infty)$, in that it contains no first-order errors, only second and higher orders. Note that all functions on the right-hand side of Eq.(15) are known at all $r$ and the variational value $\delta_v(\infty)$ can be calculated explicitly. The first order terms have been eliminated by explicit construction \cite{ref16}. The variational result in Eq.(15) provides an immediate way of improving on the JWKB approximation if $u_{JWKB}$ is chosen for the trial function. 

There is a natural physical explanation for the amplitude function being an adjoint to the phase function. In such an initial value first-order differential function for the phaseshift as Eq.(12), instead of the asymptotic phaseshift in Eq.(15), one could also ask for the value of $\delta^{(s)} (r)$ at some finite $r$. Starting again with zero value at the origin, in integrating outwards for the phaseshift that builds up as more and more of $v^{(s)}(r)$ is covered, one can write a variational expression similar to Eq.(15) but only up to that finite $r$. However, the part of $v^{(s)}(r)$ not yet ``sensed" in the interval $(r, \infty)$ would play no part were it not for the adjoint function $L(r)$ that brings that information, it or $\alpha^{-2}(r)$ with their integral running from $\infty$ to $r$. This was noted \cite{ref17} for a closely related problem of a PAM equation for $\tan \delta^{(s)} (r)$ instead of $\delta^{(s)} (r)$ itself which obeys a nonlinear Riccati equation. The same features apply, the companion amplitude function playing the role of the adjoint Lagrange function and carrying in knowledge of $v^{(s)}(r)$ from $\infty$ to complement the knowledge of it from zero to $r$. 

These features relate also to a general mathematical principle called ``invariant imbedding" which views problems such as the phaseshift not as solutions of the linear second-order Eq.(1) with twin boundary conditions at the origin and infinity but as part of a family of phaseshifts associated with the family of $v^{(s)}(r)$ that coincide with the given potential but truncated to zero beyond that value of $r$ \cite{ref17, ref18}. Such a view leads to initial value problems with just one boundary condition, at one end, but first-order differential equations albeit nonlinear. Principles of such invariant imbedding of a given problem in an extended family of similar problems are a powerful alternative in many areas of physics to the conventional approach \cite{ref18, ref19}. 

\section{Viewing PAM as gauge fixing, analogous to electromagnetism}

The key equation Eq.(7) can be cast as 

\begin{equation}
[(\frac{d}{dr} +\frac{i}{\alpha^2(r)})^2 + k^2(r)]\alpha(r) =0. 
%\label{eqn9}
\end{equation}
In the squared operator above, the linear derivative $d/dr$ term does not appear! Before proceeding to viewing Eq.(16) in terms of gauge transformations, it is worth noting that the structure of $[(d/dr + k(r)]^2 u(r)$ simplifying in form happens only for a few cases of $k(r)$ and $u(r)$. Besides the one shown above when $u = \alpha, k= i/\alpha^2$, another of interest is when $k=1/r$ and it reduces just to derivative terms $[d^2/dr^2 + (2/r)d/dr]u$. This is the basis of defining the ``radial momentum" operator $(d/dr +1/r)$ associated with just the radial kinetic energy. It may not be widely appreciated but any other multiplicative coefficient $c$ in the $1/r$ term will result in the square having other terms $(c^2-c)u$ not involving derivatives of $u$.

Rewriting Eq.(16) as       

\begin{equation}
[-\frac{1}{2}(\frac{d}{dr} +\frac{i}{\alpha^2(r)})^2 + V(r)]\alpha(r) = E\alpha(r), 
%\label{eqn9}
\end{equation}
it has the suggestive form

\begin{equation}
[-\frac{1}{2}(\frac{d}{dx_i})^2 + V(x_i)]\psi = E\psi. 
%\label{eqn9}
\end{equation}
This equation is invariant under the {\it local} gauge transformation

\begin{equation}
\psi(x_i) = \tilde{\psi}(x_i) e^{i\theta(x_i)}, 
%\label{eqn9}
\end{equation}
with $\theta$ a function of local coordinates $x_i$ (not a constant as would be in a global phase transformation). The transformed wave function satisfies 

 \begin{equation}
[-\frac{1}{2}(\frac{d}{dx_i} +iA(x_i))^2 + V(x_i)]\tilde{\psi} = E\tilde{\psi}, 
%\label{eqn9}
\end{equation}
with a ``vector'' potential $A(x_i )= \partial \theta /\partial x_i$ of a new field which simultaneously undergoes a companion gauge transformation to Eq.(19), $A_i \rightarrow A_i + i \partial \theta/\partial x_i$. 

The coupled gauge transformations are most familiar in physics in coupling a charged particle to an electromagnetic field's vector potential \cite{ref20}. The gauge arbitrariness of $\vec{A}$ to an additive gradient that was already familiar in classical electromagnetism acquired a new meaning in quantum physics, already in non-relativistic quantum mechanics, with even more significance in quantum field theory, that the wave function or field function undergoes simultaneously a phase change to preserve the physics \cite{ref21}. The electron charge appears both in the phase $\theta$ in Eq.(19) and in the $A$ in the canonical momentum operator in Eq.(20) and it is this local gauge invariance that is connected to the conservation of electric charge. This point of view of any complex quantum field requiring an accompanying vector field for gauge invariance has become central to quantum field theories today \cite{ref22}.

With this perspective, the PAM procedure can be viewed as fixing the gauge that enters naturally in separating into an amplitude and a phase. The transformations

\begin{equation}
\alpha(r)\! \rightarrow \alpha (r) e^{i\int^r \beta(r') dr'},  [1/\alpha^2(r)] \!\rightarrow [1/\alpha^2(r)] \!-\beta(r),
\end{equation}
are the counterpart of $\psi \rightarrow \psi e^{i\theta}, \vec{A} \rightarrow \vec{A}-\vec{\nabla} \theta$. In Eq.(20), the choice $\beta=0$ gives the Milne-Young-Wheeler Eq.(7). The choice $\beta = 1/\alpha^2$ gives instead the Schr\"{o}dinger Eq.(1). All choices can be capsuled by a variant of Eq.(16), valid for all $\beta$,

\begin{equation}
[(\frac{d}{dr} +\frac{i}{\alpha^2(r)}-i\beta(r))^2 + k^2(r)]\alpha(r) \exp(i\int^r \beta(r')dr') = 0. 
%\label{eqn9}
\end{equation}

\end{document}